# Examining the Examiners:
# Students' Privacy and Security Perceptions of Online Proctoring Services*

David G. Balash‡, Dongkun Kim‡, Darika Shaibekova‡
Rahel A. Fainchtein§, Micah Sherr§, and Adam J. Aviv‡
‡ *The George Washington University,* § *Georgetown University*

## Abstract

In response to the Covid-19 pandemic, educational institutions quickly transitioned to remote learning. The problem of how to perform student assessment in an online environment has become increasingly relevant, leading many institutions and educators to turn to online proctoring services to administer remote exams. These services employ various student monitoring methods to curb cheating, including restricted ("lockdown") browser modes, video/screen monitoring, local network traffic analysis, and eye tracking. In this paper, we explore the security and privacy *perceptions* of the student test-takers being proctored. We analyze user reviews of proctoring services' browser extensions and subsequently perform an online survey ($n = 102$). Our findings indicate that participants are concerned about both the amount and the personal nature of the information shared with the exam proctoring companies. However, many participants also recognize a trade-off between pandemic safety concerns and the arguably invasive means by which proctoring services ensure exam integrity. Our findings also suggest that institutional power dynamics and students' trust in their institutions may dissuade students' opposition to remote proctoring.

## 1 Introduction

In the past decade colleges and universities have steadily expanded online course offerings [19]. The Covid-19 pandemic has significantly accelerated that pace, as in-person classes were quickly replaced with virtual instruction [2]. With the increase in online education, academic integrity issues surrounding how students complete online exams led many educators to utilize *remote proctoring services* [7].[1] A 2020 EDUCAUSE poll found that more than half of higher education institutions use remote proctoring services and another 23% are either planning for or considering their use [9].

Remote proctoring services are offered by a number of companies, including popular vendors such as Respondus [29], Proctorio [25], and ProctorU [26]. Many remote proctoring services require students to install a browser extension that "locks down" their browser, preventing navigation to other sites during exam time. However, more invasive monitoring may also include webcams, screen sharing, the use of a live (human) proctor, and even automated monitoring techniques such as eye tracking and network traffic analysis.

There is evidence of higher rates of academic integrity violations for online exams [18, 22], and some argue that online proctoring is an effective tool to curb cheating [14]. However, this can come at the expense of increased test anxiety and diminished student performance [6]. Importantly, the privacy policies and practices of these services, and of online education, generally, significantly impact students and their privacy rights [4]. While concerns over the privacy and the ethics of online exam proctoring have led several institutions (cf. [1, 17]) to discontinue their contracts with online proctoring services, there is little research on the privacy perceptions and understandings *of the student test-takers* who undergo remote proctoring. In this paper we endeavor to answer the following research questions about privacy in the setting of online proctoring services:

**RQ1** What are students' perceptions and understandings of online proctoring services?
**RQ2** What are students' privacy concerns regarding the use of online proctoring software?
**RQ3** What are students' security concerns regarding the use of online proctoring software?

We first reviewed eight online proctoring services' Chrome browser extensions. Based on the number of user reviews, we observed explosive growth of online proctoring since the start of the Covid-19 pandemic (720 %). Qualitative analysis of the user reviews revealed a number of privacy concerns, including providing personal identifiable information to verify

---

*This is an extended version of a paper that appears at the *17th Symposium on Usable Privacy and Security* (SOUP'21).

[1]Remote proctoring services are sometimes called online proctoring services, or more simply, online proctoring. We use these terms interchangeably in this paper.

students' identities, live-proctors viewing webcams, local network monitoring, and screen sharing.

We subsequently developed an online survey to further explore privacy issues with $n = 102$ student participants who took an online proctored exam. Only 39 % of participants *agreed* or *strongly agreed* that they prefer an online proctored exam, and participants expressed many of the same privacy concerns as found in user reviews, particularly around the process of identity verification. They also expressed concern for installing proctoring software. A little more than half were at least *somewhat*, *moderately* or *extremely* concerned about installing proctoring software on their personal computers, and 52 % *agreed* or *strongly agreed* that exam proctoring was too privacy invasive. Participants were more comfortable with lockdown browsers, keyboard restrictions and even a live proctor while being monitored but expressed discomfort with screen, webcam or microphone recording. They were least comfortable with browser history monitoring.

Despite concerns, many participants noted a privacy-benefit trade-off in their qualitative responses, recognizing that taking exams online was more convenient and safe during the Covid-19 pandemic. At the same time, many participants also indicated that they did not believe online proctoring prevents academic dishonesty: 61 % noted that they *agreed* or *strongly agreed* that they could still cheat (if they wanted to).

We also found that power dynamics shaped students' perceptions of online proctoring. Students reported that for 97 % of remotely proctored exams, the proctoring was required by their instructor or institution. The obligatory monitoring and its backing by academic institutions may explain why many participants are able to contextualize their privacy exposure. Some participants noted that their trust in the proctoring services was due in part to their belief that their institution would not harm their security or privacy.

Given our findings, we present a number of recommendations for educators. These include acknowledging students' concerns regarding remote proctoring services, better communicating the privacy and security implications of using these services, presenting a clear rationale for using the selected proctoring system, providing some form of consent and notice to students before online proctored exams, and providing clear instructions and/or assistance in removing invasive monitoring software following an exam.

## 2 Background and Related Work

Online exam proctoring services enable students to complete an exam (or other coursework) online while being proctored remotely. When taking an online exam, students may be required to install software to assist in confirming their identity, monitoring their behavior, and preventing their access to unauthorized resources. Monitoring may include the use of the webcam and microphone, sharing computer screens, monitoring the network, eye tracking, or other behavioral tracking. Some services use a live (human) proctor to observe the student. While there are other mechanisms for remote examination, such as taking an exam using video conferencing (e.g., Zoom), this study is focused on remote online proctoring services that provide a more comprehensive observation using browser plugins and/or standalone software, as well as student identify verification. Herein, when we refer to "online exam proctoring" or an "online proctored exam" we are specifically referring to services as described above.

Despite considerable media attention [11, 13, 23, 31], student *perceptions* of online proctoring services have been understudied. We identified one recent study by Kharbat and Abu Daabes which finds high levels of privacy concerns in the UAE when using online proctoring systems [16]; we find similar results. However, unlike Kharbat and Abu Daabes, we focus on participants' security and privacy concerns, and how they compare with the risks we identified through our own analysis of these tools. A recent manuscript by Cohney et al. explores privacy risks of online proctoring services [4] but instead focuses on perceptions of university administrators and faculty; we focus on the student perspective.

Cheating during online exams has been investigated, with sometimes contradictory findings. Watson and Sottile found that students indicated they would be 4x more likely to cheat in online classes, but more readily during in-person exams [32]. Lanier compared rates of academic dishonesty at a university that offered both online and in-person learning, finding more cheating in the online courses [18]. However, Grijalva et al. found that the rate of cheating in online classes resembles that of traditionally proctored exams [10].

Hylton et al. examined whether webcam-based monitoring had a deterrent effect [14]. They found no statistically significant difference in exam scores between students who were and were not monitored, but report that non-proctored students took longer to complete exams and perceived they had more opportunity to cheat. Similarly, Rios and Liu also found little difference in exam performance on low-stakes exams, suggesting that rates of cheating are also similar between low-stakes proctored and non-proctored exams [30].

In contrast, Daffin and Jones found that student performance was generally 10-20% higher on online psychology exams that did not use online proctoring services [6]. Goedl and Malla also found that student performance was significantly greater without online proctoring. However, like Hylton et al., they found students consistently took less time to complete their exams when proctored [8]. In these studies, it is unclear whether the differences in completion time and performance were due to (1) the online proctoring acting as a deterrent to curb cheating that would otherwise result in higher test scores or (2) a psychological effect of the presence of the remote monitoring. Woldeab and Brothen addressed this more specifically and found that for students with trait test anxiety, exam-time stress was more closely correlated with poorer performance in online proctored exams [33].

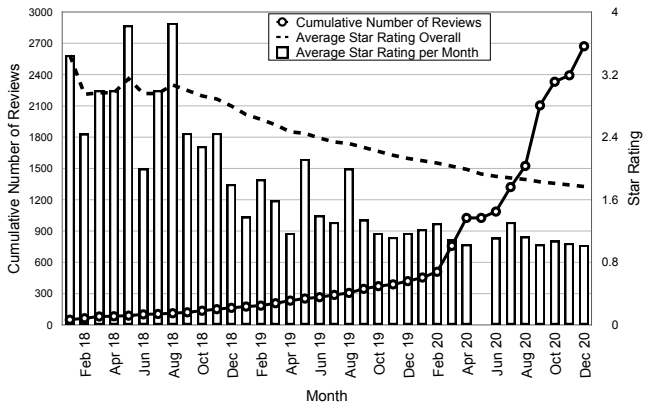

Figure 1: Number of Chrome Web Store reviews and star ratings for exam proctoring browser extensions ($n = 8$).

In two recent opinion articles, Coghlan et al. highlight ethical considerations when integrating machine learning and artificial intelligence techniques into online proctoring services [3], and Swauger opines that the algorithms that underpin online monitoring have been shown to "[reinforce] white supremacy, sexism, ableism, and transphobia" and that proctoring services inherit these traits [31]. These types of concerns, particularly those of online proctors' inadequate accessibility and protection of student privacy, have led to the cancellation or discontinuation of contracts with online proctoring vendors at the University of Illinois [17] and the University of California, Berkeley [1].

## 3 Browser Extension and Privacy Policies

As an initial investigation into online proctoring services, we conducted a study of Chrome Web Store reviews of online proctoring services' browser extensions. These extensions are often required to be installed as a prerequisite to taking online proctored exams. We analyzed user reviews from the Chrome Web Store posted between October 2015 and December 2020 for eight browser extensions. We also analyzed the privacy policies of 25 proctoring services, as reported on their websites with respect to the kinds of information collection and monitoring practices. This analysis informs the development of the online survey discussed in Section 4.

**Growth in Online Proctoring** We first analyzed the number of reviews over time, dating back to January 2018. (See Figure 1.) While there is a steady rise in the number of reviews, starting in January 2020 (the beginning of the Covid-19 pandemic) the growth in reviews greatly increased. By the end of 2020, exam proctoring browser extensions experienced an 8.2x (720 %) increase in the number of reviews, totaling 2,348 reviews. In the prior two years (2018, 2019), only 292 reviews appeared on the web store, strongly suggesting [Permission access matrix table for PSI Online, ProctorU, Proctorio, ProctorExam, Mercer Mettl, IRIS, Honorlock, ConductExam across columns: Active Tab, All Urls, Browsing Data, Clipbrd. Read, Clipbrd. Write, Content Settings, Context Menus, Cookies, Desktop Capture, Downloads, Geolocation, History, Management, Native Messaging, Notifications, Power, Privacy, Proxy, Storage, System CPU, System Display, System Mem., System Storage, Tab Capture, Tabs, Text-to-Speech, Unlmtd. Storage, Web Nav., Web Req., Web Req. Blocking] that the Covid-19 pandemic has led to a large expansion of students who are taking remotely proctored exams. This confirms a recent poll by Grajek that found that more than half of colleges and universities make use of online proctoring [9].

Table 1: The number of results found for each URL match pattern in the Honorlock browser extension manifest file. To obtain this data we used the Google site operator (e.g., `site:http://*/courses/*/quizzes/*`) in February 2021.

| Pattern | Matching URLs |
| --- | --- |
| `http://*/courses/*/quizzes/*` | 8 |
| `https://*/courses/*/quizzes/*` | 99,300 |
| `http://*/courses/*/quizzes/*/take?user_id=*` | 8 |
| `https://*/courses/*/quizzes/*/take?user_id=*` | 8 |
| `https://*/courses/*/quizzes*` | 9 |
| `*://*/d2l/lms/quizzing/*` | 8 |
| `*://*/webapps/assessment/*` | 316,000 |
| `*://*/ultra/courses/*` | 9 |
| `http://*/courses/*/quizzes` | 183,000 |
| `https://*/courses/*/quizzes` | 240,000 |
| `http://*/courses/*/quizzes/*/take` | 9 |
| `http://*/courses/*/quizzes/*/take/questions/*` | 8 |
| `https://*/courses/*/quizzes/*/take` | 231,000 |
| `https://*/courses/*/quizzes/*/take/questions/*` | 8 |
| `*://*/webapps/assessment/*` | 316,000 |
| `*://*/d2l/lms/quizzing/*` | 8 |
| Total Matches | 1,385,383 |

Table 2: Permission access of browser extensions.

Interestingly, as shown in Figure 1, there is a noticeable decline in the average star rating that coincides with the start of the pandemic (and the growth in popularity of online proctoring services). Remarkably, by the end of 2020, the average rating fell to just 1.02 (the lowest possible rating is 1).

**Analysis of Reviews** We analyzed a total of 613 reviews that were written between August 2015 and October 2020 for the browser extensions offered by ConductExam [5], Honorlock [12], IRIS [15], Mercer Mettl [21], ProctorExam [24], Proctorio [25], ProctorU [26] and PSI Online [27]. A primary coder crafted a codebook by coding a random sample of (up to) 100 reviews per extension. (Some extensions had fewer than 100 reviews.) Using the codebook, a secondary coder coded all reviews over several rounds, providing feedback on the codebook and iterating with the primary coder until inter-coder agreement was reached (Cohen's κ > 0.7).

We find that 83% (*n* = 510) of users shared negative reviews. For instance, a user stated, "Just an absolute nightmare to use," and, "It is a small wonder how they convinced all these companies to use it for their online exams." The most prevalent concern was the 55% (*n* = 335) who mentioned concerns about their privacy. For example, "I'm not letting some random person have control over facial recognition of me and scan the inside of my home." A number of reviews noted positive experiences (*n* = 73; 12%), e.g., "It has gotten the job done, and I have never had any problems with it." A few reviews (*n* = 60; 10%) mentioned that the use of proctoring services was required by their institution.

**Monitoring Techniques and Scope** We also extracted *manifest* files from each extension, which describe the permissions (or access level) for the extension and on which web pages the extension is active. Pages on which the extension is active are indicated by a list of URL match patterns, with * indicating a wild card. We found that the extensions' URL matching can be quite broad. For example, in Table 1, we report the number of Google search results that match each URL pattern specified by Honorlock's browser extension. These URL patterns match a wide variety of URLs, most likely associated with online course content hosted through Blackboard[2] or Canvas.[3] However, generic URL patterns can match other URLs (e.g., any URL that has /courses/ followed by /quizzes/), activating the browser extension regardless of whether the student is taking an exam.

This can be problematic for student privacy, beyond the duration of the exam, as these browser extensions request many browser permissions in order to conduct monitoring. Table 2 reports the permission requests for the eight proctoring browser extensions. All but two extensions request multiple permissions. ProctorU and Proctorio request the most, with Proctorio requesting 22 different permissions, which could be active when visiting any page matching a URL pattern.

**Privacy Policies** In addition to viewing permissions in the manifest file, we also reviewed the privacy policies of 25 exam proctoring services. See Question **Q6** for a full list; not all had browser extensions in the web store. Figure 2 presents the number of exam proctoring services (x-axis) that disclose certain data collection practices (y-axis). All 25 discuss setting cookies, collecting IP addresses, and accessing the webcam, and all but one note access to a photo ID to verify identity. Many policies also mention that the software will request access to the microphone, screen recordings, or collect other kinds of biometric information. Notably, 18 state that they share information with third parties.

---
[2] https://www.blackboard.com
[3] https://www.instructure.com/canvas

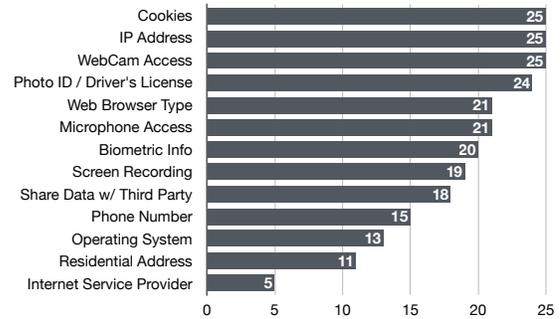

Figure 2: Data collection disclosed by exam proctoring services in their privacy policies (*n* = 25).

## 4 Survey Methodology

We conducted an online survey to evaluate the security and privacy concerns of student test-takers who are remotely proctored. The design of our study is informed by our preliminary analysis of the browser extensions and the privacy policies (see Section 3), and in what follows, we describe the survey's procedures, recruitment, limitations, and ethics. Survey results are presented in Section 5.

### 4.1 Study Procedure

To ensure that participants had taken at least one online proctored exam, we used a two-part structure with an initial *screening survey* in which qualified participants were then asked to participate in the *main study*. The full text of the screening survey and main study can be respectively found in Appendices A.1 and A.2.

**Screening Survey** We used the following two inclusion criteria to screen participants for the main study: (1) the participant is familiar with online exam proctoring and (2) the participant has taken an online proctored exam.

In the screening survey we also asked participants to describe their overall experience taking online proctored exams and to provide demographic information such as age, identified gender, education, and technical background. Participants also answered the Internet Users' Information Privacy Concerns (IUIPC) questionnaire [20] to provide insights into their privacy concerns.

**Main Study** The main study consisted of the following:
1. Informed Consent: Participants were asked to consent to the study. The consent included that participants would answer questions about their awareness and concerns about online exam proctoring services.
2. Awareness and Exposure: Participants were asked to report their experiences with online exam proctoring, including the number of exams taken, the nature of the exams, the proctoring service(s) used, and if they were required to take the exam. Participants were also asked if the online proctoring service provided any necessary accommodations or other modifications based on their needs as a test taker, and if they experienced any technical difficulties during the exam. These questions were informed by the browser extension reviews. Questions: **Q1**-**Q17**.

3. Proctoring Methods: Next, participants were asked about their level of comfort with specific monitoring methods used by proctoring services, such as eye movement tracking, video monitoring, and internet activity monitoring, and if these monitoring methods were necessary. The list of these methods were informed by the analysis of the browser extensions and privacy policies. Questions: **Q17**-**Q28**.

4. Proctoring Effectiveness: To determine the perceived effectiveness of online exam proctoring we asked if participants were less likely to cheat and if they believed it is still possible to cheat on an exam even with the monitoring methods employed by online exam proctoring services. Additionally, participants were asked if they had been accused of cheating by exam proctoring software and, if so, which specific methods such as eye movement tracking, screen recording, or internet activity monitoring was used to detect cheating. Participants could choose to not answer these questions. Questions: **Q29**-**Q33**.

5. Privacy Concerns: Participants were asked to evaluate their concern regarding sharing information with online exam proctoring companies, whether the proctoring service was a reasonable trade-off between personal privacy and the integrity of the exam, and whether online exam proctoring was a good solution for monitoring remote examinations. Questions: **Q34**-**Q39**.

6. Proctoring Software: Finally, participants were asked about the installation of exam proctoring software, what the software did, and their level of concern about the software. Questions: **Q40**-**Q50**.

### 4.2 Recruitment and Demographics

We initially recruited 27 participants by posting an advertisement on Reddit via the subreddit *SampleSize*[4] between November 14, 2020 and December 2, 2020. Note that participants recruited via Reddit did not take the screening survey, but rather the pre-survey questions were included in the main study. We excluded responses that did not meet the screening criteria.

We were not able to find a sufficiently large sample on Reddit, and so we recruited additional participants on *Prolific*[5] between December 18, 2020 and December 28, 2020. As a part of the screening survey we recruited 150 participants. Using their ProlificIDs, we re-recruited 75 of the participants who met the criteria for participation in the main study.

[4] https://www.reddit.com/r/SampleSize
[5] https://www.prolific.co

Table 3: Demographic and IUIPC data collected at the end of the screening survey.

|   |   | Screening (n = 178) | | Main Study (n = 102) | |
|---|---|---|---|---|---|
|   |   | n | % | n | % |
| Gender | Woman | 85 | 48 | 47 | 46 |
|   | Man | 85 | 48 | 52 | 51 |
|   | Non-binary | 7 | 4 | 2 | 2 |
|   | No answer | 1 | 1 | 1 | 1 |
| Age | 18–24 | 124 | 70 | 73 | 72 |
|   | 25–34 | 39 | 22 | 22 | 22 |
|   | 35–44 | 9 | 5 | 5 | 5 |
|   | 45–54 | 4 | 2 | 2 | 2 |
|   | 55+ | 2 | 1 | 0 | 0 |
|   |   | Avg. | SD | Avg. | SD |
| IUIPC | Control | 5.9 | 0.8 | 6.0 | 0.8 |
|   | Awareness | 6.5 | 0.6 | 6.5 | 0.6 |
|   | Collection | 5.7 | 1.0 | 5.7 | 1.1 |
|   | IUIPC Combined | 6.0 | 0.6 | 6.0 | 0.7 |

Participants who completed the Reddit survey were given the opportunity to enter a drawing for a $50 USD Amazon gift card with a 1 in 27 chance of winning. On average, it took 27.2 minutes (SD=24.5) to complete the Reddit survey. Participants who completed the screening survey received $0.50 USD. On average, it took 4.2 minutes (SD=2.7) to complete the screening survey and 15 minutes (SD=7.1) to complete the main study. Participants who completed the main study received $3.50 USD.

Seventy-two percent of main study participants were between 18–24 years old, 22 % were between 25–34 years old, and 7 % were 35 years or older. The identified gender distribution for the main study was 51 % men, 46 % women, and 3 % non-binary or did not disclose gender. Participant characteristics are presented in Table 3, and additional demographic information can be found in Appendix B. In total, $n = 102$ participants were recruited for the main study.

### 4.3 Ethical Considerations and Limitations

The study protocol was approved by our Institutional Review Board (IRB) with approval number `REDACTED`, and all collected data is associated with random identifiers. Throughout this process we considered that many participants may not want to share their perceptions of whether proctoring services are effective at preventing academic dishonesty, and so we made those questions optional.

Our study is limited in its recruitment, particularly to Prolific and Reddit users residing in the U.S. We cannot claim full generalizability of the results. Despite this limitation, prior work [28] suggests that online studies about privacy and security behavior can approximate behaviors of populations.

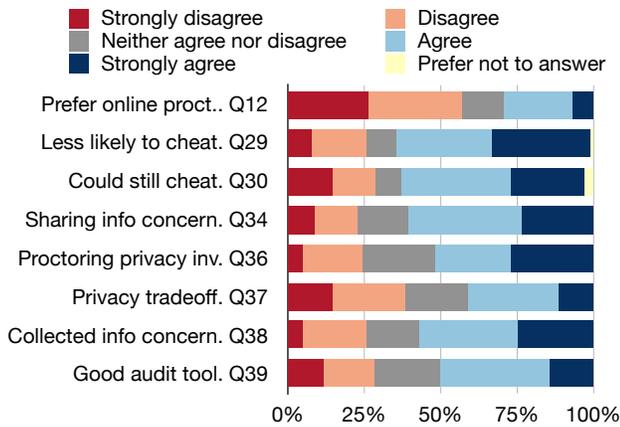

Figure 3: Impressions of online proctoring services.

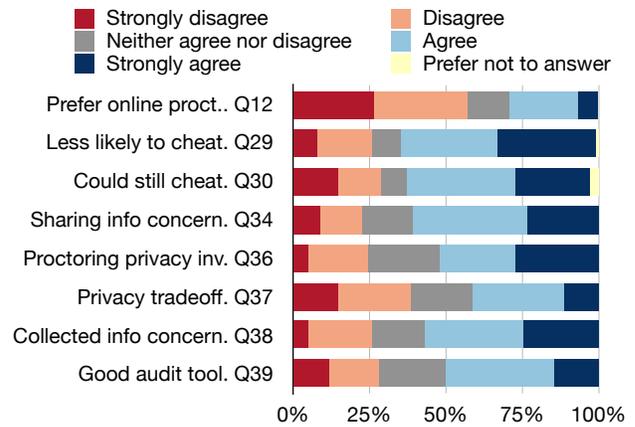

Figure 4: Encountered exam requirements.

We are also limited by the fact that this study relies on self-reported behavior. We cannot verify that the participants actually experienced an online proctored exam, which is why we used a screening survey. Finally, responses can suffer from social desirability and response bias, leading participants to over describe their awareness of online exam proctoring as they may believe that this is the expectation of the researchers. Such biases may be most present when participants indicate concerns and indicate they are less likely to cheat on an exam.

## 5 Results

We organize our results according to our research questions. We first present our findings concerning participants' perceptions and understanding of online exam proctoring (**RQ1**), and then describe participants' privacy concerns regarding online exam proctoring (**RQ2**). Finally, we discuss participants' understanding of exam proctoring software and their concerns about such software (**RQ3**).

For all qualitative findings, we used a pair of primary coders from the research team, each of whom crafted a codebook and identified descriptive themes by coding each question. A secondary coder coded a 20 % sub-sample from each of the free-response questions over several rounds, providing feedback on the codebook and iterating with the primary coder until inter-coder agreement was reached (Cohen's $\kappa > 0.7$).

### 5.1 RQ1: Perceptions and Understanding

As part of **RQ1**, we seek to measure (1) student perceptions of online proctoring and (2) their understanding of the methods used by online proctoring services to monitor exams.

**Experience with Exam Proctoring** Nearly half ($n = 49$; 48 %) of respondents had taken five or more online proctored exams, 38 % ($n = 39$) had taken between two to four (inclusive), and a mere 14 % ($n = 14$) of participants had only taken a single online-proctored test (**Q1**). The online proctoring service *Respondus* was the most used ($n = 20$; 20 %), followed by *Proctorio* ($n = 13$; 13 %), and *ProctorU* ($n = 10$; 10 %) (**Q6**) (see Figure 10 in Appendix B). This generally conforms to the survey conducted by EDUCAUSE [9]. The most common exam proctoring methods used to monitor study participants included: lockdown browser ($n = 71$; 70 %), webcam recording ($n = 65$; 64 %), screen recording ($n = 61$; 60 %), live proctor ($n = 60$; 60 %), and microphone recording ($n = 51$; 50 %) (**Q23**). (See Figure 5.)

While most participants ($n = 94$; 92 %) reported that at least one of their online proctored exams was a course exam (e. g., test, midterm exam, final exam), many ($n = 47$; 46 %) had also used online exam proctoring for lower stakes course assessments such as quizzes (**Q2**). The most common subjects that were proctored included science ($n = 24$; 24 %), business ($n = 17$; 17 %), mathematics ($n = 16$; 16 %), computer science ($n = 11$; 11 %), and medicine ($n = 9$; 9 %).

Many of the participants ($n = 83$; 81 %) took their most recent online proctored exam in the year 2020 during Covid-19, and most were in the last half of 2020: December ($n = 39$; 47 %), November, ($n = 16$; 19 %), and October ($n = 8$; 10 %) (**Q4**). This matches the explosive growth in browser reviews described in Section 3.

**Requirement to Use Online Proctoring** Nearly all participants were required to use an online proctoring service: 97 % of subjects ($n = 99$) noted they had been required to take an exam using online proctoring services (**Q7**) by an authority at their university. When asked who had required them to take an online proctored exam (**Q8**), 70 % ($n = 68$) of respondents indicated their class instructor, followed by 23 % ($n = 22$) who reported that online proctoring was required by their university. Only 7 % ($n = 7$) of participants indicated their requirement to use online proctoring had stemmed from having taken a standardized test.

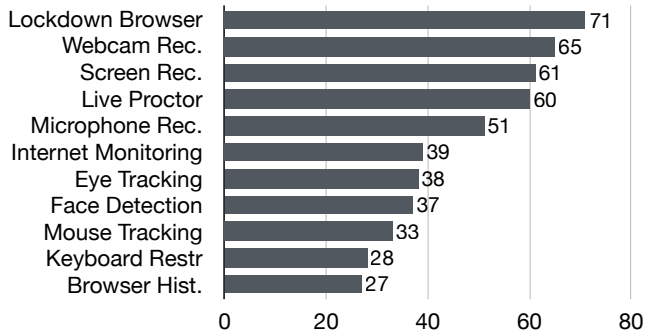

Figure 5: Prevalence of monitoring types (**Q23**).

**Preference for/against Online Proctoring** A majority of participants ($n = 58$; 56 %) prefer traditional exam formats, but others ($n = 30$; 30 %) preferred online proctored exams (**Q12**; Figure 3). Still, half ($n = 51$; 50 %) stated that they *agree* ($n = 36$; 35 %) or *strongly agree* ($n = 15$; 15 %) that online exam proctoring is a good solution for monitoring remote exams (**Q39**). We asked participants to qualitatively explain some of the benefits of using online exam proctoring (**Q10**): 42 % ($n = 43$) highlighted that they prevent cheating, e.g., "It effectively prevents cheating so students abilities can be graded accurately" (P51); and 29 % ($n = 30$) liked taking exams remotely, e.g., "You don't have to leave your house to take the exam" (P102). Some participants ($n = 5$; 5 %) specifically mentioned the social distancing during the Covid-19 pandemic, e.g., "It was a good way to still be able to take exams securely while distance learning because of COVID-19" (P35). Other participants ($n = 12$; 12 %) liked the flexibility, e.g., "It is a bit nicer to be able to take the exam at a different time that works best for me" (P3).

To explore the factors that may drive a preference for or against taking an online proctored exam, we performed an ordinal logistic regression. For the outcome variable, we used the Likert response to **Q12**, preference for online exam proctoring over traditional exam formats. The factors we considered were participant responses to questions about the number of exams taken, awareness of monitoring methods, concern about the amount of information collected, general privacy perceptions, privacy trade-off, online exams as a good solution, discomfort with monitoring methods, and concern about sharing information. Each of the considered factors was converted to a binary variable, using the appropriate Likert values as bins. Table 6 in Appendix B presents the full regression table.

We find that those that *agree* or *strongly agree* that online proctoring is a good solution for remote examination were 3.66x more likely to have a higher preference for online exams ($b = 1.30, OR = 3.66, p = 0.01$). A lack of privacy concerns also played a role: those that either *disagree* or *strongly disagree* that online proctored exams are privacy invasive were at a significantly increased likelihood of preferring online proctored exams ($b = 2.21, OR = 9.10, p < 0.001$). Surprisingly, if participants *disagree* or *strongly disagree* that they are concerned about the amount of information being collected, they are 5.8x less likely to prefer online exams ($b = -1.76, OR = 0.17, p = 0.03$). At the same time, participants who noted that they are *uncomfortable* or *very uncomfortable* with observation methods during exams were 2.6x less likely to prefer online exams ($b = -0.95, OR = 0.39. p = 0.05$).

The above suggests that while privacy concerns play a role in students' preference for online proctoring, concerns about data collection may not resonate as a privacy concern. Instead, concern about monitoring methods, as we discuss in Section 5.2, appear to be of higher consequence for participants.

**Preventing Cheating** Online exam proctoring is perceived as a deterrent to cheating. When asked if online exam proctoring makes it less likely for them to cheat, 63 % ($n = 65$) of participants agreed or strongly agreed, while only 26 % ($n = 26$) disagreed or strongly disagreed (**Q29**). However, 60 % ($n = 61$) agreed or strongly agreed that it would still be possible for them to cheat during an online proctored exam, with only 29 % ($n = 29$) who disagreed or strongly disagreed (**Q30**).

When asked to qualitatively explain their belief about the ability to cheat, 21 % ($n = 21$) responded that a second device such as a smartphone could be used, and 13 % ($n = 13$) reported that notes, cheat sheets, or other materials could be used to cheat. Others ($n = 17$; 17 %) explained that it was difficult to cheat. For example, P93 said, "I think that with so many sources being monitored on the student's end, this would make it extremely difficult for them to cheat." Only 2 % of participants ($n = 2$) reported being accused of cheating by the exam proctoring software (**Q32**).

**Experiences with Monitoring** When asked to described their overall experience being monitored during their exam (**Q25**), some participants ($n = 26$; 25 %) reported that being monitored was a negative experience. For example, P62 responded, "I felt uncomfortable because I do not like being watched," and P64 stated,

> ... it felt much more stressful than ... taking an exam in a typical proctored environment. I feared that any little movement or sound may trigger the system and flag me for cheating...

For a minority of participants ($n = 18$; 18 %), being monitored was a positive experience. For instance, P50 stated, "It was pretty good, I stayed focused on the test."

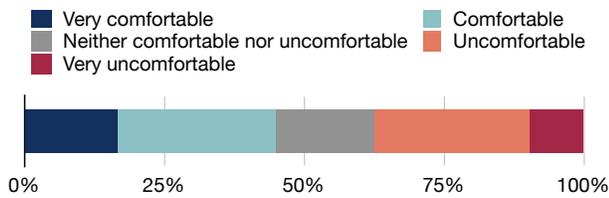

Figure 6: General comfort with proctoring methods (**Q22**).

However, other participants ($n = 15$; 15 %) had privacy concerns about being monitored, including P27 who shared, "It does feel uncomfortable to have my person and screen recorded via video, knowing that the recordings are saved for at least some period of time," and P55 who noted, "Its [sic] terribly intrusive and not worth the possibility that students will cheat."

For some participants ($n = 12$; 12 %), being monitored was a distraction or caused increased stress that was detrimental to their exam performance. For example, P66 indicated, "It creates a very stressful environment that prevents me from working to the best of my abilities," and P22 described it as "icky and uncomfortable" and that they felt like they "had to perform in a certain way because I didn't know if someone was watching."

**RQ1 Key Findings** Many students took an online-proctored exam in the wake of the Covid-19 pandemic, which corresponds to our analysis of browser extension reviews (Section 3). Participants predominantly did not take online proctored exams by choice but were rather required to do so by their instructors. By and large, participants have taken multiple exams with a remote proctor. At the same time, most respondents would prefer a traditional exam even while acknowledging that online exam proctoring is a good solution for remote exams. Those who think online proctoring is a good solution for remote examination as well as those who do not think proctored exams are privacy invasive are more likely to prefer online exam proctoring. We found that concern about data collection matters less than concern about monitoring methods when it comes to privacy and exam preference. Participants also largely believed that exam proctoring deters cheating, but most felt that it was still possible to cheat, particularly using a second device.

## 5.2 RQ2: Privacy Concerns

We next investigate students' privacy concerns regarding online exam proctoring (**RQ2**).

**Comfort with Monitoring Methods** When asked about their general comfort level with the methods used to proctor their exam (**Q22**), participants were slightly more comfortable overall (see Figure 6): 45 % ($n = 46$) were either *comfortable* or *very comfortable* with monitoring, while 37 % ($n = 38$) were either *uncomfortable* or *very uncomfortable*. (The remaining participants ($n = 18$; 18 %) were *neither comfortable nor uncomfortable*.)

We also asked participants about their comfort with specific monitoring methods (**Q28**). Aggregated results are presented in Figure 7. To compare the comfort across monitoring methods, we additionally performed a Kruskal-Wallace H-test ($H = 94.6, p < 0.001$) which showed significant difference, and a post-hoc, pair-wise Mann-Whitney U test (with Holm-Sidek correction) indicated that those differences are dominant when comparing monitoring via lockdown browser (participants' most comfortable monitoring method) and all other methods, except for live proctoring and keyboard restrictions. (See Table 7 in Appendix B.) In particular, there are significant differences with some of the most common monitoring methods: webcam recording, screen recordings, and microphone recordings. This suggests that some of the methods deemed most invasive are among those that are used most often, and this in turn may drive students' privacy concerns.

To explore the factors affecting monitoring comfort further, we performed an ordinal logistic regression with an outcome variable of the Likert response to **Q22** (overall comfort with exam privacy) to reported comfort with individual proctoring methods, binning *comfortable* and *very comfortable*. The full regression table (Table 5) appears in Appendix B. We find that comfort with live proctoring ($b = 1.20, OR = 3.31, p < 0.001$) and webcam recordings ($b = 1.96, OR = 7.08, p < 0.001$) significantly increased the likelihood of being more comfortable with exam proctoring generally, suggesting that discomfort with these forms of observation is problematic for many students; both were commonly experienced, cf. Figure 5.

Participants were also asked how necessary a given monitoring method is for online proctoring (see Figure 8). Again there is a significant difference (Kruskal-Wallace: $H = 92.9, p < 0.001$), and a post-hoc analysis (see Table 8 in Appendix B) revealed that there are significant differences between lockdown browser (deemed most necessary) and live proctoring, microphone recording, browser history monitoring, keyboard restrictions, eye tracking and mouse tracking. There were no differences between the trio of lockdown browser, webcam and screen recording with respect to how necessary they are perceived to be for online proctoring. Webcam and screen recording were not (pair-wise) significantly different than live proctoring.

**Sharing Information** Part of the process of taking an online proctored exam is to verify the identity of the exam taker. This process may involve proof of identification via physical documentation such as IDs and other forms of identity checks that may require students to provide sensitive information to online proctoring services. Students may also be required to create accounts on these services to facilitate that process, and we find that 44 % ($n = 45$) of study par-

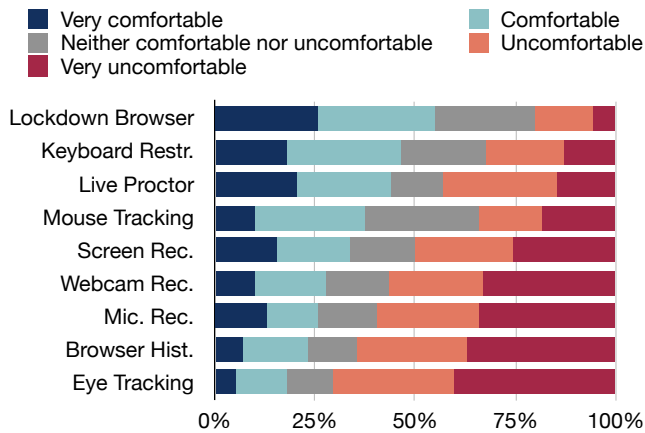

Figure 7: Comfort with monitoring types (**Q28**).

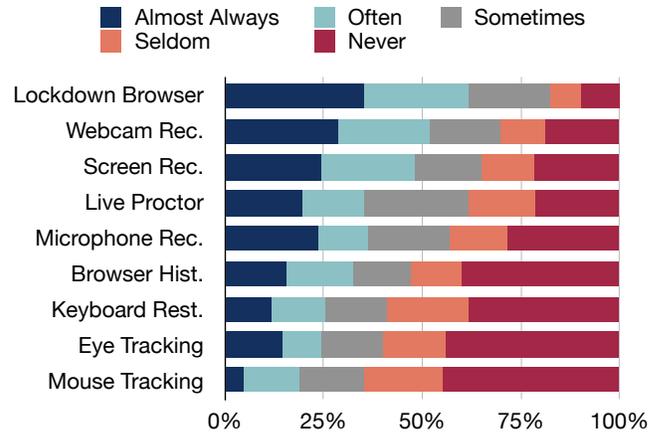

Figure 8: Necessity of monitoring types (**Q27**).

ticipants were required to do just that (**Q18**). Participants also reported that many forms of personal information were required during account creation and before taking an exam, such as full name ($n = 56$; 55 %), student ID number ($n = 52$; 51 %), email address ($n = 51$; 50 %), educational institution ($n = 39$; 38 %), birth date ($n = 29$; 28 %), phone number ($n = 19$; 19 %), residential address ($n = 16$; 16 %), driver's licence number ($n = 10$; 10 %), and social security number ($n = 7$; 7 %) (**Q19**; see Figure 11 in Appendix B). For some participants, physical documentation was required; these included student IDs ($n = 56$; 55 %), driver's licenses ($n = 32$; 31 %), and passports ($n = 7$; 7 %) (**Q20**; see Figure 12 in Appendix B). When asked if they were concerned about sharing this kind of information with online exam proctoring companies, most participants ($n = 62$; 61 %) *agreed* ($n = 38$; 37 %) or *strongly agreed* ($n = 24$; 24 %) (**Q34**). Of those who responded with concerns (**Q35**), being uncomfortable sharing personal information was the most common explanation ($n = 28$; 27 %). For instance, P91 shared, "I feel uneasy that in order to take an exam, I have to share personal information," and P45 said, "I understand that if I opt to take a test online it needs to be fairly taken, but that doesn't mean I should open up these proctoring companies up to my home…"

Data collection was also a concern for some participants ($n = 17$; 17 %). For example P101 responded, "I am not sure what they will do with my information and how long they will store/keep my information," and P58 shared, "For things like recording my computer, or accessing my browser history, I feel like that could invite abuse that go beyond simply making sure I'm honestly taking an exam…" Other participants ($n = 28$; 27 %) had no concerns about sharing information with exam proctoring services, such as P53, who said, "I'm not anymore [sic] worried about it than I am sharing my info with the school," and P48, who said, "I feel since it was required by my school it is a safe place to share information."

**Privacy Trade-off** When asked if they thought online exam proctoring was too privacy invasive, 52 % ($n = 53$) of study participants *agreed* ($n = 25$; 25 %) or *strongly agreed* ($n = 28$; 27 %) that it was too privacy invasive (**Q36**). There was a split between those who agreed that online exam proctoring offered a reasonable trade-off between personal privacy and exam integrity and those who disagreed (**Q37**). Forty-one percent ($n = 42$) of participants *agreed* ($n = 30$; 29 %) or *strongly agreed* ($n = 21$; 21 %) while 39 % ($n = 39$) of participants *disagreed* ($n = 24$; 24 %) or *strongly disagreed* ($n = 15$; 15 %). We also find evidence of split opinions regarding online proctoring in the qualitative results. In response to **Q11**, $n = 11$ (11 %) of participants reported that there was a trade-off between privacy and academic integrity. For example, P41 noted, "I think it is a valid reason to use online exam proctoring…during this time pandemic. …I can understand giving up some privacy to ensure integrity of exam results."

Participants also reported being concerned about the *amount* of information that online proctoring services collect during the exam (**Q38**). Fifty-seven percent ($n = 58$) of participants *agreed* ($n = 33$; 32 %) or *strongly agreed* ($n = 25$; 25 %), while 26 % ($n = 26$) of participants *disagreed* ($n = 21$; 21 %) or *strongly disagreed* ($n = 5$; 5 %). We again see similar results in qualitative responses in **Q11**: 59 % ($n = 60$) reported a privacy concern, with concerns about webcam access being the most common ($n = 27$; 26 %). For example, P65 reported, "I believe that online exams can be invasive, as at least mine required both a webcam and microphone, so they could see me and my room and hear my surroundings," and P36 responded:

> …*Unlike in-class proctoring, students must be filmed in their homes…The view is also on the student* 100 % *of the time so the student cannot relax and has their entire body language and quirks on display. It is a breach of privacy without enough benefit to justify it.*

Some participants ($n = 6$; 6 %) had concerns about relinquishing control of their computing devices to the exam proc-toring services, e.g., "It is a little scary about how much they can access and control your device" (P101). Sharing of personal information was a concern for participants ($n = 6$; 6 %), such as P39, who noted, "It does make me a little uncomfortable that there is a 3rd party company that may have my personal identification and see into my room." Still other participants ($n = 19$; 19 %) reported that they had no privacy concerns, such as P69, who said, "I don't see any huge issues with privacy in online exam proctoring," and P51 who stated:

> *I don't mind that they can see my room and control my screen. They aren't doing anything sinister, and I can revoke all permissions at the end of the exam.*

**RQ2 Key Findings** A majority of students found online exam proctoring to be privacy invasive, most citing concerns with the webcam and microphone recordings, which provides the means to view, listen, and record inside a student's room. However, some students felt a trade-off between loss of personal privacy and exam integrity was reasonable.

When considering privacy in the context of preventing academic dishonesty, participants had mixed reactions to the proctoring methods used. Lockdown browsers, webcams, and screen recordings were viewed as necessary for online exam proctoring compared to other methods, and these were also the most commonly used methods to observe students during an exam. However, only about a quarter of the respondents were comfortable with webcam or screen recording, while half were comfortable with lockdown browsers. Regression analysis indicated that comfort with live proctoring and webcam recordings drive comfort overall with monitoring. This suggests that there is a gap between the proctoring methods commonly used and the comfort level of students with those observation techniques.

Students also create accounts and share significant information with online proctoring services to verify their identities. Services often require personal information, such as a student ID number, email address, phone number, and residential address, as well as images of physical documentation such as student IDs and driver's licenses. Participants expressed concern about sharing this and other personal information with exam proctoring companies. They were also concerned about the overall quantity of information collected, what would be done with the information, and how long it would be stored.

## 5.3 RQ3: Security Concerns

Along with privacy concerns, the installation of specialized software to enable proctoring could lead to security issues, and as part of addressing **RQ3**, we surveyed participants about their experiences and concerns with proctoring software.

**Browser Extensions** One way in which exam proctoring companies provide monitoring during an exam is by requiring students to install web browser extensions, as noted in

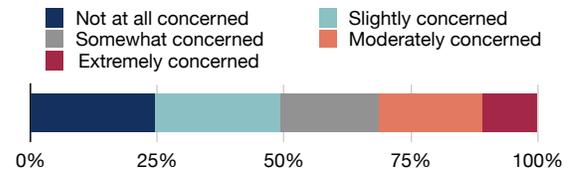

Figure 9: Concern over installing proctoring software (**Q49**).

Section 3. Most study participants ($n = 65$; 64 %) were required to install a web browser extension to take their exam (**Q40**). When we asked participants who installed a browser extension what they thought the extension did (**Q41**), they responded that the extension locked down their browser ($n = 27$; 42 %), collected data ($n = 8$; 12 %), monitored network activity ($n = 8$; 12 %), initiated screen recording ($n = 7$; 11 %), disabled functionality on their device ($n = 6$; 9 %), and enabled their webcam ($n = 6$; 9 %) and microphone ($n = 4$; 6 %).

Proctorio ($n = 13$; 13 %) was the most common web browser extension installed by study participants, followed by ProctorU ($n = 13$; 13 %), and Honorlock ($n = 13$; 13 %) (**Q42**; Figure 13 in Appendix B). Significantly, only 45 % ($n = 31$) of participants who installed a web browser extension reported removing or disabling the extension after completing their exam (**Q43**). Given that many of these browser extensions have pervasive monitoring permissions that can be activated on a broad set of URLs (see Section 3), it is important that students remove these extensions; the failure of 45 % ($n = 30$) to do so suggests that installing this custom software may put students at risk beyond the exam.

**Standalone Software** Another way in which exam proctoring companies provide monitoring during the examination is by requiring students to install standalone software, which we define as software that is installed as an application on their computer and is not a browser extension. Thirty-five percent ($n = 36$) of participants reported they were required to install exam proctoring software (not including a browser extension) (**Q44**). When we asked participants who installed exam proctoring software what they thought the software did (**Q45**), they responded that the proctoring software locked down their browser ($n = 11$; 31 %), disabled functionality on their device ($n = 8$; 22 %), monitored their activity ($n = 7$; 19 %), initiated screen recording ($n = 3$; 8 %), and enabled their webcam ($n = 3$; 8 %). Of the participants who installed exam software, most ($n = 32$; 89 %) said that they did uninstall the exam proctoring software after the exam was complete (**Q44**). Only one participant reported having issues uninstalling the exam proctoring software (**Q44**). Most participants ($n = 88$; 86 %) said they had installed the exam software on their personal computer (**Q44**). When asked to report their concern for installing proctoring software (Figure 9), 52 % of participants ($n = 52$) specified that they were at least *somewhat* concerned ($n = 20$; 20%), *moderately concerned* ($n = 21$; 21%), or even *extremely concerned* ($n = 11$; 11%). In contrast, 48% ($n = 50$) of participants were *slightly concerned* ($n = 25$; 24%) or *not at all concerned* ($n = 25$; 24%) (**Q49**).

We asked participants to explain their concern, or lack of concern, regarding the installation of online exam software (**Q49**). Many participants ($n = 49$; 48%) explained that they had privacy concerns. Some ($n = 11$; 11%) replied that the software's potential access to sensitive personal information was a concern; for instance, P32 said, "I am worried it will be able to access sensitive information," and P38 reported, "It is my own computer which stores all of my information so that is a bit iffy." A few participants ($n = 6$; 6%) had concerns about data collection from their computer after the exam was completed, such as P52, who said, "I wonder if they continued collecting information after the exam," and P102, who noted, "I don't trust software that is designed to gather information about my activities to confine itself to being used only for exams." Others ($n = 11$; 11%) were unsure what information could be collected by the software; e.g., P69 and P34, who respectively stated "It's unclear what all data it's collecting and when it's running," and "I don't know what information it was collecting or how it would be used." Still others ($n = 28$; 27%) had no concerns about the software. A few ($n = 5$; 5%) stated they had no concerns because the exam privacy software was supported by their university. For instance, P40 noted, "I believe the university would not use the proctoring service if their software was dangerous," and P87 stated, "I know that my school and professors wouldn't have me install anything that could harm my computer or invade my privacy."

**RQ3 Key Findings** The browser extension is the most common way in which exam proctoring tools access students computing devices. Students understand that these extensions are used both to surveil them and their devices during the exam and to disable functionality that would otherwise allow them to access unauthorized resources. Despite this knowledge, only a small number of students actually removed or disabled the extension after completing their exam, leaving permission-hungry software residing on their computers.

Standalone software is also used for exam proctoring, and most students install this software on their own personal computers. Students are concerned about this software and say they worry that it may access personal information stored on their computers. Most students did uninstall this standalone software, an action that highlights and confirms their concerns.

## 6 Recommendations and Conclusions

**Privacy Trade-offs During a Pandemic** Given the necessary rapid transition to remote learning, many institutions and educators did not have sufficient time to restructure courses around alternative forms of learning and skills assessment. Content and exams that had originally been envisioned as in-person suddenly had to be delivered and proctored remotely, forcing institutions to seek solutions to a perceived exam integrity problem. Our results suggest that students understand that their educational institutions were struggling to maintain mandated safety protocols while continuing to provide academic rigor, as evidenced by the fact that a large number of study respondents (41%) reported that online exam proctoring was a reasonable trade-off between personal privacy and exam integrity. A recurring theme in the qualitative responses was that giving up some personal privacy during the Covid-19 pandemic to maintain safety protocols was a valid reason to accept the use of online proctoring services. This suggests that student acceptance of online exam monitoring is higher than it might otherwise be in a post-pandemic situation.

At the same time, we find that a large percentage of students have significant concerns—e.g., sharing personal information with proctoring companies, the amount of information collected by these companies, and installing online exam proctoring software on their computers. When we consider these facts, it is clear that many students found their proctored exams to be privacy invasive and would prefer alternatives to online proctored exams. However, it is unclear if a post-pandemic context will lead to increased student opposition to invasive monitoring or if students will have become accustomed to these proctoring tools and resigned to their use.

*Recommendation:* Based on this study, we recommend that institutions and instructors both expand student choice by developing alternative forms of student assessment that can account for privacy concerns whenever practicable and plan to reduce future reliance on exam proctoring services after the Covid-19 pandemic.

**Necessary Type of Monitoring** Many participants agreed that the ability to deter cheating during an exam is important, up to a point, after which they felt that monitoring goes from necessary to unnecessary and invasive. In fact, we find that the types of monitoring that students perceive as the most unnecessary are among those that they report are the most uncomfortable and invasive. For instance, a majority of students reported that they do not think it is necessary to monitor mouse movement, eye movement, or web browser history; correspondingly, a majority also reported being uncomfortable with the monitoring of their eye movement, web browser history, microphone, and webcam. These very monitoring types are those that students refer to most when they discuss how they feel that the online proctored exam can create a stressful environment, how they feel "watched," and how they worry that any small sound or tiny movement—such as looking away from the screen briefly—could flag them for cheating. Students report that these additional stressors and anxieties distract them, reduce their focus, and prevent them from performing to the best of their abilities. This level of monitoring assumes cheating and pre-penalizes all students with additional stress and anxiety whether they were planning to be honest or dishonest.

Our work suggests that even though technologically advanced invigilation techniques are available, such as 360-degree room scans and eye movement tracking, it does not mean that they are either necessary to curb cheating or sensitive to students' personal privacy and device security. Continuing to use monitoring techniques that students find unnecessary for exam integrity, while at the same time requiring students to sacrifice their personal privacy, displays a lack of trust for students and undermines students' trust in educators and institutions.

*Recommendation:* We recommend that institutions and educators follow a principle of least monitoring by using the minimum number of monitoring types necessary, given the class size and knowledge of expected student behavior. Institutions should perform due diligence when selecting online exam proctoring companies with whom to contract, and they should take into account student privacy, student discomfort for certain monitoring types, and software installation requirements. Moreover, instructors should use caution when selecting monitoring types while setting up exams and should provide students with clear reasoning for having selected the individual methods that will be used to monitor their exams.

**Invasion of Personal Computers**  As we have seen, exam proctoring browser extensions and standalone software contain invasive monitoring tools. These tools often include permissions to access the webcam, microphone, and web browser history. However, 43 % of students did not remove or disable the required browser extensions once they had completed their assessments. As is the case with any custom software, there is risk of vulnerabilities in these extensions. The fact that students often neglect to remove them therefore creates increased potential for harm from loss of privacy or security intrusions. Institutions should therefore be sensitive to which proctoring software they require students to install on their personal devices.

*Recommendation:* We recommend that institutions thoroughly review common vulnerabilities and exposures of the online exam proctoring software they plan to license for installation on students' personal computers. We would also recommend that institutions limit the installation of standalone exam proctoring software to devices issued to students by the institution.

**Implied Trust via Institutional Support**  Exam proctoring tools are often integrated with existing learning management software, such as Blackboard and Canvas, giving the appearance that they are a part of the standard educational software stack and imparting a sense of safety and normalcy. At the same time, institutions have spent large amounts of money to obtain site license agreements for exam proctoring software. This gives the appearance of a certain amount of due diligence being applied to the purchase, while potentially increasing the barriers to student resistance towards these forms of examination.

Throughout our qualitative findings are statements from students of a transfer of trust between institutions who licence and faculty who support the exam proctoring software and the software itself. These students say that they believe their university would not use the software to proctor exams if it was dangerous. Moreover, they believe that their school would not have them install anything that could harm their computer or invade their privacy. Institutional support for third-party proctoring software, which conveys credibility, makes the exam proctoring software appear safer and less potentially problematic because students assume that institutions have done proper vetting of both the software and the methods employed by the proctoring services.

*Recommendation:* We recommend that the students, along with faculty and administrators, take part in the assessment and selection of exam proctoring software. Students should be involved in every step of the process, from deciding whether to use exam proctoring software to determining which, if any, software should be used and which methods should be made available for exam monitoring.

**Power Imbalances**  Finally, when students' options are limited to taking an online proctored exam or failing the course, it is a clear indication of an institutional power dynamic; 97 % of students indicated they were required to take an online proctored exam. Offering students more choices for assessment and being upfront with students about institutional privacy norms is a crucial step to alleviate this power imbalance.

*Recommendation:* We recommend implementing notice and choice for courses employing online exam proctoring and allowing students to consent to any monitoring that will take place during course quizzes and exams. We also recommend that syllabi include a readable privacy policy to better communicate expectations.

## Acknowledgements

This work is partially funded by the National Science Foundation under grants 1718498 and 1845300, and the Georgetown University Callahan Family Professor of Computer Science Chair Fund.

# A Survey Instruments

## A.1 Screening Survey

**S1** How familiar are you with online exam proctoring?
- ○ Not at all familiar
- ○ Slightly familiar
- ○ Somewhat familiar
- ○ Moderately familiar
- ○ Extremely familiar

**S2** I have taken an online proctored exam.
- ○ Yes
- ○ No
- ○ Unsure
- ○ Prefer not to answer

**S3** Please describe your overall online proctored exam experience.
Answer: ____________

*These questions were followed by the 10 IUIPC items as described by Malhotra et al. [20]*

**D1** What is your gender?
- ○ Woman
- ○ Man
- ○ Non-binary
- ○ Prefer not to disclose
- ○ Prefer to self-describe

**D2** What is your age?
- ○ 18 – 24
- ○ 25 – 34
- ○ 35 – 44
- ○ 45 – 54
- ○ 55 – 64
- ○ 65 or older
- ○ Prefer not to disclose

**D3** Are you currently a student?
- ○ Yes
- ○ No
- ○ Prefer not to disclose

**D4** What is the highest degree or level of school you have completed?
- ○ No schooling completed
- ○ Some high school, no diploma
- ○ High school graduate, diploma, or equivalent (e. g., GED, Abitur, baccalaureat)
- ○ Some college credit, no degree
- ○ Trade / technical / vocational training
- ○ Associate degree
- ○ Bachelor's degree
- ○ Master's degree
- ○ Professional degree (e. g., J.D., M.D.)
- ○ Doctorate degree
- ○ Prefer not to disclose

**D5** Which of the following best describes your educational background or job field?
- ○ I have an education in, or work in, the field of computer science, computer engineering or IT.
- ○ I do not have an education in, nor do I work in, the field of computer science, computer engineering or IT.
- ○ Prefer not to disclose

## A.2 Main Study

Thank you for participating in the second part of our survey. You have been invited to our main study because of your direct experience taking an online proctored exam.

Your answers based on your online exam proctoring experiences are important to us!

**Please read the following instructions carefully:**
- Take your time in reading and answering the questions.
- Answer the questions as accurately as possible.
- It is okay to say that you don't know an answer.

**Q1** How many online proctored exams have you taken?
- ○ 1
- ○ 2
- ○ 3
- ○ 4
- ○ 5+

**Q2** What was the nature of the exam(s) you took using an online proctoring service? Select all that apply.
- ○ Course Quiz
- ○ Course Exam (E.g. test, midterm exam, final exam)
- ○ Standardized Test (E.g. GRE, GMAT, bar exam)
- ○ I have not taken an exam with online proctoring
- ○ Other: ____________

**Q3** Of those ones you chose, which is the most recent?
- ○ Course Quiz
- ○ Course Exam (E.g. test, midterm exam, final exam)
- ○ Standardized Test (E.g. GRE, GMAT, bar exam)
- ○ I have not taken an exam with online proctoring
- ○ *[Other value entered in Q2]*

**Q4** As best as you can remember, when was the month, date, and year when you last took an online exam with a proctoring service?
- ○ Month: ____________
- ○ Day: ____________
- ○ Year: ____________

**Q5** What was the subject matter of the last examination you took using an online proctoring service?
Answer: ____________

**Q6** What was the name of the online proctoring service used during your last examination?
- ○ ConductExam
- ○ Pearson OnVUE
- ○ PSI Online Proctoring
- ○ Examity
- ○ ProctorExam
- ○ Questionmark
- ○ ExamSoft
- ○ ProctorFree
- ○ Respondus
- ○ Honorlock
- ○ Proctorio
- ○ Smowl
- ○ IRIS Invigilation
- ○ Proctortrack
- ○ Surpass
- ○ Kryterion
- ○ ProctorU
- ○ Talview
- ○ Mercer Mettl
- ○ Proview
- ○ TestReach
- ○ Other: ____________
- ○ Unsure

**Q7** Were you required to take that exam using an online exam proctoring service?
- ○ Yes, I was required to use an online exam proctoring service.
- ○ No, there were other forms of assessment available to me but I opted to use an online exam proctoring service

**Q8** Who required you to take an online proctored exam?
Answer: ____________

**Q9** What were the deciding factors in your choice to take your exam with an online proctoring service instead of other forms of assessment?
Answer: ____________

**Q10** In your experience, what are some benefits of using online exam proctoring?
Answer: ____________

**Q11** Please explain your views on the privacy of online exam proctoring.
Answer: ____________

**Q12** I prefer online exam proctoring services over traditional exam formats.
- ○ Strongly disagree
- ○ Disagree
- ○ Neither agree nor disagree
- ○ Agree
- ○ Strongly agree

**Q13** Did the online proctoring service make any necessary exam accommodations or other modifications based on your needs as an exam taker?
- ○ Yes, I request and was provided adequate accommodations
- ○ No, I requested and was not provided adequate accommodations
- ○ I did not request nor require exam accommodations
- ○ Unsure
- ○ Prefer not to answer

**Q14** Please describe the accommodations provided to you. You may indicate N/A if you prefer not to answer. *[Shown only if answer to Q13 was "Yes"]*
Answer: ________________

**Q15** Please describe how accommodations were not provided for you despite your request. You may indicate N/A if your prefer not to answer. *[Shown only if answer to Q13 was "No"]*
Answer: ________________

**Q16** Did you experience any technical difficulties when taking your exam as it relates to the online proctored service?
○ Yes ○ No ○ Unsure

**Q17** Please explain any technical difficulties you may have experienced during your exam with the online proctored service. *[Shown only if answer to Q15 was "Yes"]*
Answer: ________________

### Online Exam Proctoring Methods
In this part of the survey you will be asked about the methods employed by online exam proctoring services.

**Q18** When preparing to take an online proctored exam were you required to create an account with the online proctoring service?
○ Yes ○ No ○ Unsure

**Q19** When registering for an online proctored exam what, if any, personal information where you required to enter in online forms? Select all that apply.
○ Residential Address
○ Educational institution affiliation
○ Email Address
○ Student ID Number
○ Full Name
○ Social Security Number
○ Driver's License Number
○ Phone Number
○ No information was required
○ Unsure
○ Other: ________

**Q20** When taking an online proctored exam what kinds of physical documentation, if any, were you required to provide? Select all that apply.

○ Driver's License
○ Student ID
○ Passport
○ No physical documentation was required
○ Unsure
○ Other: ________

**Q21** How aware are you of the methods used by online exam proctoring services to monitor exam takers?
○ Not at all aware ○ Moderately aware
○ Slightly aware ○ Extremely aware
○ Somewhat aware

**Q22** How comfortable were you with the methods used to proctor the exam(s) that was proctored online?
○ Very comfortable
○ Comfortable
○ Neither comfortable nor uncomfortable
○ Uncomfortable
○ Very uncomfortable

**Q23** Please select all methods that were used to proctor the exam(s) that was proctored online. Select all that apply.
○ Live proctor visible to me
○ Live proctor not visible to me
○ Web browser history monitoring
○ Eye movement tracking
○ Facial detection
○ Lockdown browser
○ Mouse movement tracking
○ Keyboard restrictions (E.g. no copy and paste)
○ Screen recording
○ Microphone recording
○ Internet activity monitoring (E.g. interaction with a web site)
○ Webcam recording

**Q24** Please describe any methods not listed above, or indicate "none" if there are no more methods.
Answer: ________________

**Q25** Please describe your overall experience being monitored during your online proctored exam.
Answer: ________________

**Q26** Previously you indicated that your most recent online proctored exam was a *[Answer from Q3]*. Please refer to that experience in answering the following questions. Considering your most recent online proctored exam, what kind of monitoring (if any) was necessary to proctor that exam online?
Answer: ________________

**Q27** Previously you indicated that your most recent online proctored exam was a *[Answer from Q3]*. Again considering your most recent exam from the question above. For each exam monitoring type please select how often they are necessary for online proctoring.

| | Always | Often | Sometimes | Rarely | Never |
|---|---|---|---|---|---|
| Live proctor | ○ | ○ | ○ | ○ | ○ |
| Web browser history monitoring | ○ | ○ | ○ | ○ | ○ |
| Eye movement tracking | ○ | ○ | ○ | ○ | ○ |
| Lockdown browser | ○ | ○ | ○ | ○ | ○ |
| Mouse movement tracking | ○ | ○ | ○ | ○ | ○ |
| Keyboard restrictions (E.g. no copy/paste) | ○ | ○ | ○ | ○ | ○ |
| Screen recording | ○ | ○ | ○ | ○ | ○ |
| Microphone recording | ○ | ○ | ○ | ○ | ○ |
| Webcam recording | ○ | ○ | ○ | ○ | ○ |

**Q28** Again considering your most recent online proctored exam of *[Answer from Q3]*. For each exam monitoring type please select how comfortable you feel about them.

| | Very Comfortable | Comfortable | Neither comfortable nor uncomfortable | Uncomfortable | Very Uncomfortable |
|---|---|---|---|---|---|
| *[Types from Q27]* | ○ | ○ | ○ | ○ | ○ |

### Online Exam Proctoring Functionality
In this part of the survey you will be asked about the functionality of online exam proctoring services. If you feel uncomfortable answering any question below, you may select "Prefer not to answer."

**Q29** The use of online exam proctoring tools makes it less likely that my classmates or I will cheat on an exam.
○ Strongly disagree ○ Agree
○ Disagree ○ Strongly agree
○ Neither agree nor disagree ○ Prefer not to answer

**Q30** If I wanted to, I believe I would still be able to cheat even with online exam proctoring.
○ Strongly disagree ○ Agree
○ Disagree ○ Strongly agree
○ Neither agree nor disagree ○ Prefer not to answer

**Q31** Please explain your belief about the ability to cheat (or not cheat) with online proctored exams. Or write N/A if you prefer not to answer.
Answer: ________________

**Q32** Have you been accused of cheating by exam proctoring software?

○ Yes    ○ Prefer not to answer
○ No

**Q33** Which of the following methods employed by the online exam proctoring software was used to accuse you of cheating? Select all that apply. *[Shown only if answer to Q32 was "Yes"]*
- ○ Live proctor visible to me
- ○ Live proctor not visible to me
- ○ Web browser history monitoring
- ○ Eye movement tracking
- ○ Facial detection
- ○ Lockdown browser
- ○ Mouse movement tracking
- ○ Keyboard restrictions (E.g. no copy and paste)
- ○ Screen recording
- ○ Microphone recording
- ○ Internet activity monitoring (E.g. interaction with a web site)
- ○ Webcam recording
- ○ Unsure
- ○ Other: ______
- ○ Prefer not to answer

**Privacy Concerns**   In this part of the survey you will be asked about the benefits and potential risks you associate with online exam proctoring.

**Q34** I am concerned about sharing information with online exam proctoring companies.
- ○ Strongly disagree    ○ Agree
- ○ Disagree    ○ Strongly agree
- ○ Neither agree nor disagree

**Q35** Please explain your answer to the previous question regarding the consequences of sharing information.
Answer: ______

**Q36** I think online exam proctoring services are too privacy invasive.
- ○ Strongly disagree    ○ Agree
- ○ Disagree    ○ Strongly agree
- ○ Neither agree nor disagree

**Q37** I think online exam proctoring offers a reasonable tradeoff between my privacy and the integrity of the exam.
- ○ Strongly disagree    ○ Agree
- ○ Disagree    ○ Strongly agree
- ○ Neither agree nor disagree

**Q38** I am concerned about the amount of information that online proctoring services collect during the exam.
- ○ Strongly disagree    ○ Agree
- ○ Disagree    ○ Strongly agree
- ○ Neither agree nor disagree

**Q39** I think online exam proctoring is a good solution for monitoring remote examinations.
- ○ Strongly disagree    ○ Agree
- ○ Disagree    ○ Strongly agree
- ○ Neither agree nor disagree

**Exam Proctoring Web Browser Extensions**   A web browser extension is a small software module that is used to extend the functionality of your web browser with additional features. Some online proctoring services require exam takers to install a web browser extension in order to take an exam. In this part of the survey you will be asked questions about your experience using web browser extensions to take online proctored exams.

**Q40** Were you required to install and use a web browser extension in order to participate in a proctored online exam?
○ Yes    ○ No    ○ Unsure

**Q41** What do you think the browser extension did? *[Shown only if answer to Q40 was "Yes"]*
Answer: ______

**Q42** What was the most recent web browser extension you installed in order to participate in a proctored online exam? *[Shown only if answer to Q40 was "Yes"]*
- ○ ConductExam    ○ ProctorExam    ○ Unsure
- ○ Examity    ○ Proctorio    ○ No browser extension was installed
- ○ Honorlock    ○ ProctorU
- ○ IRIS Invigilation    ○ PSI Online Proctoring    ○ Other: ______
- ○ Mercer Mettl

**Q43** Did you remove or disable any browser extensions that you were required to install to take an online proctored exam? *[Shown only if answer to Q40 was "Yes"]*
○ Yes    ○ No    ○ Unsure

**Exam Proctoring Software**   Some online proctoring services require exam takers to install standalone application software on a computer, like your PC or Mac, in order to take an exam. In this part of the survey you will be asked questions about your experience installing and using exam application software to take online proctored exams. Please note this may be in addition to the requirement to install a browser extension.

**Q44** Did you have to install other types of exam proctoring software (not including a browser extension)?
○ Yes    ○ No    ○ Unsure

**Q45** What do you think this exam proctoring software did? *[Shown only if answer to Q44 was "Yes"]*
Answer: ______

**Q46** Did you uninstall the exam proctoring software? *[Shown only if answer to Q44 was "Yes"]*
○ Yes    ○ No    ○ Unsure

**Q47** Did you have any issues uninstalling the exam proctoring software? *[Shown only if answer to Q46 was "Yes"]*
○ Yes    ○ No    ○ Unsure

**Q48** From the computing devices listed below, please select the device you used to take your most recent online proctored exam.
- ○ Personal Computer    ○ Mobile Device (Smartphone, Tablet, etc)
- ○ Shared Home Computer
- ○ School Issued Computer    ○ Unsure
- ○ Public Computer (E.g. Library)    ○ Other: ______

**Q49** How concerned are you about installing online exam proctoring software on the computer you used to take the exam?
- ○ Not at all concerned    ○ Moderately concerned
- ○ Slightly concerned    ○ Extremely concerned
- ○ Somewhat concerned

**Q50** Please explain your answer to the previous question regarding your concern about installing online exam proctoring software.
Answer: ______

# B   Additional Figures and Tables

| Metric | Reddit sample | Prolific sample |
|---|---|---|
| Total Participants | 27 | 75 |
| Gender: Man | 10 (37%) | 42 (56%) |
| Gender: Woman | 14 (52%) | 33 (44%) |
| Gender: Nonbinary | 2 (7.4%) | 0 |
| Gender: Prefer not to disclose | 1 (3.7%) | 0 |
| Age: 18-24 | 17 (63%) | 56 (75%) |
| Age: 25-34 | 8 (30%) | 14 (19%) |
| Age: 35-44 | 2 (7.4%) | 3 (4.0%) |
| Age: 45-54 | 0 | 2 (2.7%) |
| Age: 55-64 | 0 | 0 |
| Student | 21 (78%) | 69 (92%) |
| NonStudent | 6 (22%) | 6 (8.0%) |
| Some high school (no diploma) | 0 | 1 (1.3%) |
| High school graduate, diploma, or equivalent | 2 (7.4%) | 9 (12%) |
| Trade / technical / vocational training | 1 (3.7%) | 0 |
| Some college credit, no degree | 5 (19%) | 35 (47%) |
| Associate's degree | 3 (11%) | 11 (15%) |
| Bachelor's degree | 14 (52%) | 15 (20%) |
| Master's degree | 1 (3.7%) | 2 (2.7%) |
| Professional degree (e.g., J.D., M.D.) | 1 (3.7%) | 1 (1.3%) |
| Schooling : Other (including PhD) | 0 | 1 (1.3%) |
| IT background | 11 (41%) | 20 (27%) |
| No IT background | 15 (56%) | 55 (73%) |
| IT background: Prefer not to disclose | 1 (3.7%) | 0 |

Table 4: Participant demographics for Reddit and Prolific participants. Prolific demographics exclude tallies from respondents who only completed the pre-survey.

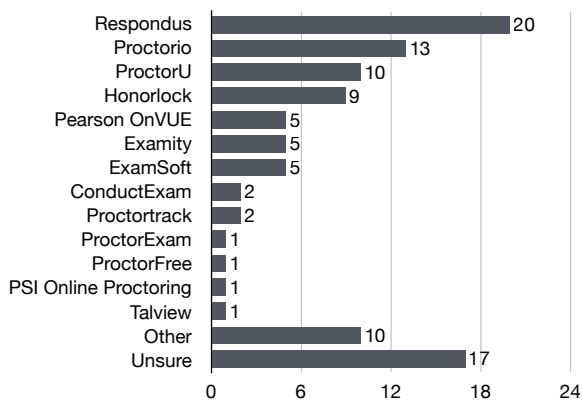

Figure 10: Proctoring services experienced. This generally conforms to the survey conducted by EDUCAUSE [9].

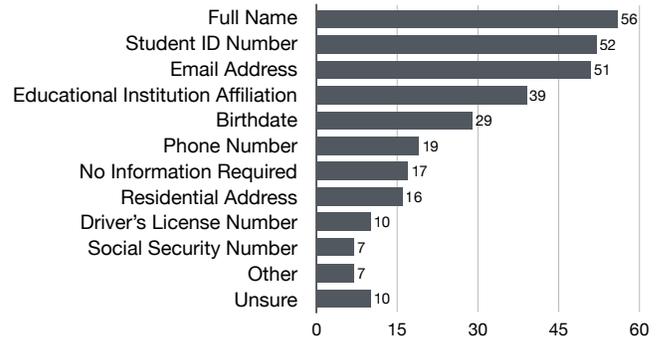

Figure 11: Information required when registering for an online proctored exam (**Q19**).

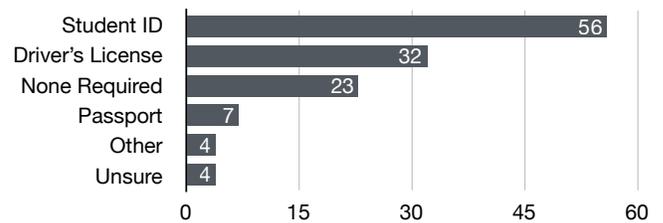

Figure 12: Physical documentation required to provide (**Q20**).

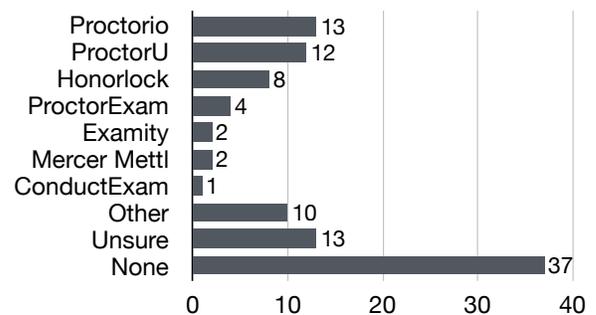

Figure 13: Most common browser extensions installed (**Q42**).

Table 5: Ordinal regression model to describe the level of comfort with proctoring methods responses to Question **Q22**. The model uses an ascending comfort scale (i. e., from *Very uncomfortable* to *Very comfortable*). The Aldrich-Nelson pseudo $R^2$ of the model is 0.58.

| Factor | Estimate | Odds ratio | Error | t value | Pr(>\|z\|) | |
|---|---|---|---|---|---|---|
| Live Proctor $\in$ {*Comfortable, Very Comfortable*} | 1.20 | 3.31 | 0.42 | 2.88 | <0.001 | ** |
| Browser History $\in$ {*Comfortable, Very Comfortable*} | −0.49 | 0.61 | 0.54 | −0.91 | 0.36 | |
| Eye Tracking $\in$ {*Comfortable, Very Comfortable*} | −0.80 | 0.45 | 0.72 | −1.11 | 0.27 | |
| Lockdown Browser $\in$ {*Comfortable, Very Comfortable*} | −0.05 | 0.95 | 0.46 | −0.11 | 0.91 | |
| Mouse Tracking $\in$ {*Comfortable, Very Comfortable*} | 0.76 | 2.13 | 0.54 | 1.40 | 0.16 | |
| Keyboard Restr. $\in$ {*Comfortable, Very Comfortable*} | −0.19 | 0.83 | 0.48 | −0.39 | 0.70 | |
| Screen Recording $\in$ {*Comfortable, Very Comfortable*} | 1.00 | 2.72 | 0.53 | 1.87 | 0.06 | . |
| Mic Recording $\in$ {*Comfortable, Very Comfortable*} | 1.11 | 3.04 | 0.67 | 1.66 | 0.10 | . |
| Webcam Recording $\in$ {*Comfortable, Very Comfortable*} | 1.96 | 7.08 | 0.63 | 3.11 | <0.001 | ** |
| **Intercepts** | | | | | | |
| *Very uncomfortable* \| *Uncomfortable* | −1.42 | 0.24 | 0.39 | −3.61 | <0.001 | *** |
| *Uncomfortable* \| *Neither comfortable nor uncomfortable* | 0.59 | 1.81 | 0.33 | 1.81 | 0.07 | . |
| *Neither comfortable nor uncomfortable* \| *Comfortable* | 1.67 | 5.32 | 0.37 | 4.52 | <0.001 | *** |
| *Comfortable* \| *Very comfortable* | 4.17 | 64.80 | 0.61 | 6.89 | <0.001 | *** |

**Signif. codes:** '***' $\hat{=}$ < 0.001; '**' $\hat{=}$ < 0.01; '*' $\hat{=}$ < 0.05; '.' $\hat{=}$ < 0.1

Table 6: Ordinal regression model to describe the preference for online proctored exams based on responses to Question **Q12**. The model uses an ascending agreement scale (i. e., from *Strongly disagree* to *Strongly agree*). The Aldrich-Nelson pseudo $R^2$ of the model is 0.75.

| Factor | Estimate | Odds ratio | Error | t value | Pr(>\|z\|) | |
|---|---|---|---|---|---|---|
| Exams taken > 3 | −0.07 | 0.93 | 0.40 | −0.18 | 0.86 | |
| Aware methods $\in$ {*Moderately aware, Extremely aware*} | 0.83 | 2.28 | 0.67 | 1.23 | 0.22 | |
| Concern amount $\in$ {*Disagree, Strongly disagree*} | −1.76 | 0.17 | 0.81 | −2.17 | 0.03 | * |
| Privacy invasive $\in$ {*Disagree, Strongly disagree*} | 2.21 | 9.10 | 0.66 | 3.35 | <0.001 | *** |
| Reasonable tradeoff $\in$ {*Agree, Strongly agree*} | 0.94 | 2.57 | 0.56 | 1.69 | 0.09 | . |
| Good solution $\in$ {*Agree, Strongly agree*} | 1.30 | 3.66 | 0.52 | 2.48 | 0.01 | * |
| Comfort methods $\in$ {*Uncomfortable, Very uncomfortable*} | −0.95 | 0.39 | 0.49 | −1.93 | 0.05 | . |
| Concern sharing $\in$ {*Disagree, Strongly disagree*} | 0.58 | 1.78 | 0.81 | 0.71 | 0.48 | |
| **Intercepts** | | | | | | |
| *Strongly disagree* \| *Disagree* | −0.67 | 0.51 | 0.50 | −1.34 | 0.18 | |
| *Disagree* \| *Neither agree nor disagree* | 1.40 | 4.07 | 0.54 | 2.61 | 0.01 | ** |
| *Neither agree nor disagree* \| *Agree* | 2.39 | 10.90 | 0.57 | 4.17 | <0.001 | *** |
| *Agree* \| *Strongly agree* | $1.14E+02$ | 0.72 | 6.61 | $3.74E-11$ | | *** |

**Signif. codes:** '***' $\hat{=}$ < 0.001; '**' $\hat{=}$ < 0.01; '*' $\hat{=}$ < 0.05; '.' $\hat{=}$ < 0.1

Table 7: Post-Hoc Analysis of comfort with monitoring method **Q28** using pair-wise Mann-Whitney U-Test with Holm-Sidek Correction.
(Kruskal Wallace: $H = 94.6, p < 0.001$)

|  | Lockdown browser | Keyboard Restr. | Live proctor | Mouse Tracking | Screen Rec. | Webcam Rec. | Mic. Rec. | Browser Hist. |
|---|---|---|---|---|---|---|---|---|
| Lockdown browser | — | | | | | | | |
| Keyboard Restr. | 0.553 | — | | | | | | |
| Live proctor | 0.232 | 0.930 | — | | | | | |
| Mouse Tracking | 0.021* | 0.822 | 0.930 | — | | | | |
| Screen Rec. | 0.001* | 0.261 | 0.666 | 0.857 | — | | | |
| Webcam Rec. | < 0.001* | 0.005* | 0.039* | 0.160 | 0.822 | — | | |
| Mic. Rec. | < 0.001* | 0.003* | 0.027* | 0.095 | 0.764 | 0.930 | — | |
| Browser Hist. | < 0.001* | < 0.001* | 0.001* | 0.007* | 0.267 | 0.920 | 0.930 | — |
| Eye Tracking | < 0.001* | < 0.001* | < 0.001* | < 0.001* | 0.032* | 0.635 | 0.764 | 0.930 |

Table 8: Post-Hoc Analysis of neccessity of monitoring method **Q27** using pair-wise Mann-Whitney U-Test with Holm-Sidek Correction.
(Kruskal Wallace: $H = 92.8, p < 0.001$)

|  | Lockdown Browser | Webcam Rec. | Screen Rec. | Live Proctor | Mic. Rec. | Browser Hist. | Keyboard Restr. | Eye Tracking |
|---|---|---|---|---|---|---|---|---|
| Lockdown Browser | — | | | | | | | |
| Webcam Rec. | 0.605 | — | | | | | | |
| Screen Rec. | 0.171 | 0.946 | — | | | | | |
| Live Proctor | 0.004* | 0.582 | 0.911 | — | | | | |
| Mic. Rec. | 0.004* | 0.490 | 0.865 | 0.946 | — | | | |
| Browser Hist. | < 0.001* | 0.010* | 0.084 | 0.472 | 0.677 | — | | |
| Keyboard Restr. | < 0.001* | < 0.001* | 0.009* | 0.092 | 0.336 | 0.946 | — | |
| Eye Tracking | < 0.001* | < 0.001* | 0.004* | 0.044* | 0.181 | 0.946 | 0.946 | — |
| Mouse Tracking | < 0.001* | < 0.001* | < 0.001* | < 0.001* | 0.010* | 0.605 | 0.865 | 0.946 |

## C Qualitative Codebooks

### C.1 Browser Extension Reviews Codebook

- **dislikes-product (510)**

- **privacy-invasive (335)**
  *personal-information (30), invasion-of-network (22), webcam-access (22), screen-recording-activity (13), data-retention (9), modifying-data (7), data-collection-without-knowledge (6), microphone-access (6), tracks-location (3),*

- **technical-issues (183)**
  *does-not-work (73), affects-device-performance (13), false-alarms (12),*

- **likes-product (73)**
  *easy-to-use (8), fast-to-install (4), saves-time (4),*

- **forced-by-institution (60)**

- **not-recommended-for-use (43)**

- **prefer-different-method-of-testing (40)**

- **difficult-to-use (40)**

- **does-not-reflect-testing-environment (31)**

- **uninstall-after (30)**

- **spyware (27)**

- **bad-product-support (24)**
  *long-wait-time (6)*

- **affects-testing-time (19)**

- **malware (15)**

- **course-grade-affected (13)**
  *failed-exam (10)*

- **mentally-affected (12)**

- **creepy (11)**

- **not-accommodating-for-students-no-device (9)**

- **skeptical-of-reviews (7)**

- **ineffective-against-cheating (4)**

- **good-product-support (4)**

- **disables-certain-computer-functions (3)**

- **bot-live-chat (2)**

### C.2 Main Study Codebook

- **privacy-concern (188)**
  *personal-information (47), webcam (36), data-collection (17), unsure-what-information-collected (14), data-collection-after-exam (9), data-retention (9), microphone (9), control-over-device (8), data-breach (6), admin-access (5), third-party-data-sharing (5), screen (3), privacy-right-violation (3), computer-activity (3), sale-of-personal-information (3), no-disclosure-provided (1), downloading-software (1), viewing-information-without-consent (1), pii-details (1), downloads (1), invasion-of-network (1), potentially-being-hacked (1)*

- **benefit (96)**
  *prevents-cheating (44), remote-location (30), schedule-flexibility (12), easy-to-use (9), comfortable (8), convenient (6), testing-during-pandemic (5), quick-grading (5), focus (4), less-stress (3), less-distraction (3), standardized-experience (2), grading-accuracy (2), saves-time (2), more-time (1), no-cost (1), reduced-resources (1), less-error-prone (1), accountability (1)*

- **necessary-to-proctoring (85)**
  *webcam (29), lockdown-browser (27), screen-recording (22), microphone (16), live-proctor (14), eye-movement (12), browsing-history (8), disable-new-tabs (7), check-test-taking-area (5), facial-tracking (4), disable-copy-paste (4), mouse-tracking (4), facial-identification (3), keyboard-tracking (2), no-access-to-notes (2), time-limit (2), watch-for-communication-with-other-students (1), watch-for-suspicious-activity (1), no-cell-phones (1), disable-short-cuts (1), flag-suspicious-activity (1), hand-visibility (1)*

- **no-concerns (77)**
  *supported-by-university (5), can-be-uninstalled (1), no-installation-required (1), basic-verification (1), trust-proctoring-services (1), do-not-cheat (1), no-personal-information-required (1)*

- **possible-to-cheat (65)**
  *using-another-device (21), cheat-sheet (13), omit-areas-of-room (2), depends-on-software (1), use-other-tabs (1), easier-than-in-person (1)*

- **lockdown-browser (39)**

- **technical-difficulties (32)**
  *connection-issues (5), long-loading-time (4), webcam (4), images-blocked-by-service (2), does-not-work (2), affects-device-performance (2), ended-test (2), falsely-flagged-as-cheating (2), lockdown-browser-pause (2), pause-because-no-face-detection (1), minor (1), exam-would-not-start (1), poor-audio-quality (1), internet-down-exam-stops*

- *(1), submission-issues (1), not-compatible-with-wireless-accessories (1), microphone (1)*
- **negative-experience (26)**
  *testing-environment (14), somebody-watching (1)*
- **uncomfortable (22)**
  *testing-environment (11), downloading-software (3)*
- **mentally-affected (20)**
  *stress (11), nervous (3), worried (3), focus (1)*
- **positive-experience (18)**
- **difficult-to-cheat (18)**
  *webcam-recording (1)*
- **neutral (16)**
- **disabled-functions (14)**
- **reasonable (12)**
- **unnecessary-to-proctoring (10)**
- **forced-by-institution (10)**
- **somewhat-concerned (10)**
- **screen-recording (10)**
- **webcam (9)**
- **unsure-what-information-collected (9)**
- **failed-to-answer-question (8)**
- **monitor-network-activity (8)**
- **data-collection (8)**
- **monitored-activity (7)**
- **prefer-different-method-of-testing (7)**
- **monitor-browser-history (6)**
- **no-issues (6)**
- **likes-service (6)**
- **no-benefits (5)**
- **facilitate-exam-proctoring (5)**
- **proctor (5)**
  *did-not-help (1), asked-to-relocate (1)*
- **spyware (5)**
- **ineffective-testing-environment (4)**
- **microphone (4)**
- **easy-to-use (4)**
- **cheating-is-bad (4)**
- **do-not-know-what-it-did (3)**
- **proctoring-software (3)**
  *admin-access (2), disabled-functions (2), screen-recording (1), lockdown-browser (1)*
- **personal-information (3)**
- **uncomfortable-downloading-software (3)**
- **webcam-monitoring (3)**
- **no-technical-difficulties (3)**
- **lack-of-trust (3)**
- **delete-extension-after-exam (3)**
- **does-not-make-sense (2)**
- **secure (2)**
- **unaware-of-privacy-issues (2)**
- **do-not-know (2)**
- **secure-data (2)**
- **not-accommodating (2)**
  *equipment (1), student-situations (1)*
- **difficult-to-use (2)**
  *long-set-up-time (1)*
- **potential-virus (2)**
- **unauthorized-credit-card-check (2)**
  *negative-impact-on-credit-score (2)*
- **privacy-resigned (2)**
- **admin-access (1)**
- **allowed-use-of-tools (1)**
- **keyboard-input (1)**
- **affected-testing-performance (1)**
  *no-extra-sheets-allowed-during-test (1)*
- **no-privacy-concern (1)**
- **proctor-was-quiet (1)**
- **bad-product-support (1)**
  *long-wait-time (1)*

- **proctored-exam (1)**

  *face-recognition (1), weird (1), no-other-choice (1)*

- **proctor-was-helpful (1)**

  *restarted-exam (1)*

- **professor (1)**

  *monitored-the-test (1)*

- **affects-quality-of-education (1)**
- **prefer-not-to-download-extensions (1)**
- **afraid-to-cheat (1)**
- **webcam-recording (1)**
- **connected-to-another-desktop (1)**
- **provided-instructions (1)**
- **checked-for-required-software (1)**
- **monitored-activity (1)**
- **hacked-computer (1)**
- **device-provided-for-proctoring (1)**
- **prevents-cheating (1)**

- **user-recommendation (1)**

  *involve-government (1)*

- **proctor-interruption-during-test (1)**
- **there-will-always-be-cheaters (1)**
- **cheating-is-subjective (1)**
- **tracked-key-strokes (1)**
- **mouse-tracking (1)**
- **proctoring-companies (1)**

  *need-to-know-how-students-feel (1)*

- **slowed-down-computer (1)**
- **could-not-search-test-questions (1)**
- **audio-recording (1)**
- **proctor-set-the-exam (1)**
- **no-download-required (1)**
- **falsely-flagged-as-cheating (1)**

  *opening-of-new-tab (1)*

- **prevented-cheating (1)**